\documentclass[angeo]{copernicus}
\begin{document}

\title{Sizes of flaring kernels in various parts of the H$\alpha$ line profile}

\author{K. Radziszewski}
\author{P. Rudawy}

\affil{Astronomical Institute of University of Wroc{\l}aw, 51-622
Wroc{\l}aw, ul. Kopernika 11, Poland}


\runningtitle{Sizes of the flaring kernels}

\runningauthor{Radziszewski {\&} Rudawy}

\correspondence{K. Radziszewski\\(radziszewski@astro.uni.wroc.pl)}

\received{}
\pubdiscuss{} 
\revised{} \accepted{} \published{}


\firstpage{1}

\maketitle

\begin{abstract}
In this paper we present new results of spectra-photometrical
investigations of the flaring kernels' sizes and their intensities
measured simultaneously in various parts of the H$\alpha$ line
profile. Our investigations were based on the very high temporal
resolution spectral-imaging observations of the solar flares
collected with Large Coronagraph (\emph{LC}), Multi-channel
Subtractive Double Pass Spectrograph and Solar Eclipse Coronal
Imaging System (\emph{MSDP-SECIS}) at Bia{\l}kow Observatory
(University of Wroc{\l}aw, Poland).

We have found that the areas of the investigated individual
flaring kernels vary in time and in wavelengths as well as the
intensities and areas of the H$\alpha$  flaring kernels decreased
systematically when observed in consecutive wavelengths toward the
wings of the H$\alpha$ line. Our result could be explained as an
effect of the cone-shaped lower parts of the magnetic loops
channeling high energy particle beams exciting chromospheric
plasma.
\end{abstract}

\introduction

Solar flares are investigated for more than a century, nevertheless
some very important questions still remain unanswered. One of them
is a problem of very fast heating of a relatively dense and cold
chromospheric plasma by charged particles streaming down along
flaring loops from primary energy sources located somewhere close to
the tops of the loops and - to some extent - by conduction and by
electromagnetic radiation. The accelerated particles travel with
nearly relativistic speeds and collide with the chromosphere in
fractions of seconds after their release.

Satellite-born observations made in hard X-ray domain of the
spectrum clearly reveal that primary regions of the magnetic energy
conversion are located high in the corona. There are also secondary
hard X-ray sources, located near feet of the flaring loops
(so-called "foot-point HXR sources"). The co-temporal observations
made in visible wavelengths (like in hydrogen H$\alpha$ line, 656.3
nm) reveal usually numerous bright compact and/or extended emission
sources localized in the closest vicinity of the foot-point HXR
sources. The high time resolution observations of the visible
flaring kernels show that their intensities undoubtedly vary in
time.

\begin{figure*}[t]
\vspace*{2mm}
\begin{center}
\includegraphics[width=13cm]{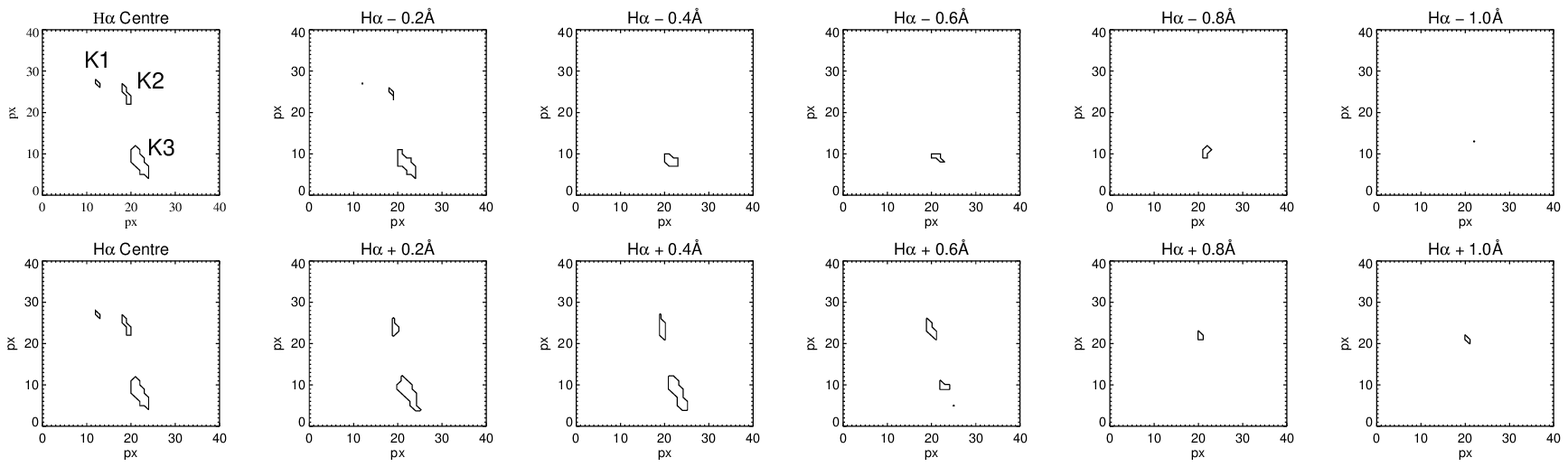}
\includegraphics[width=13cm]{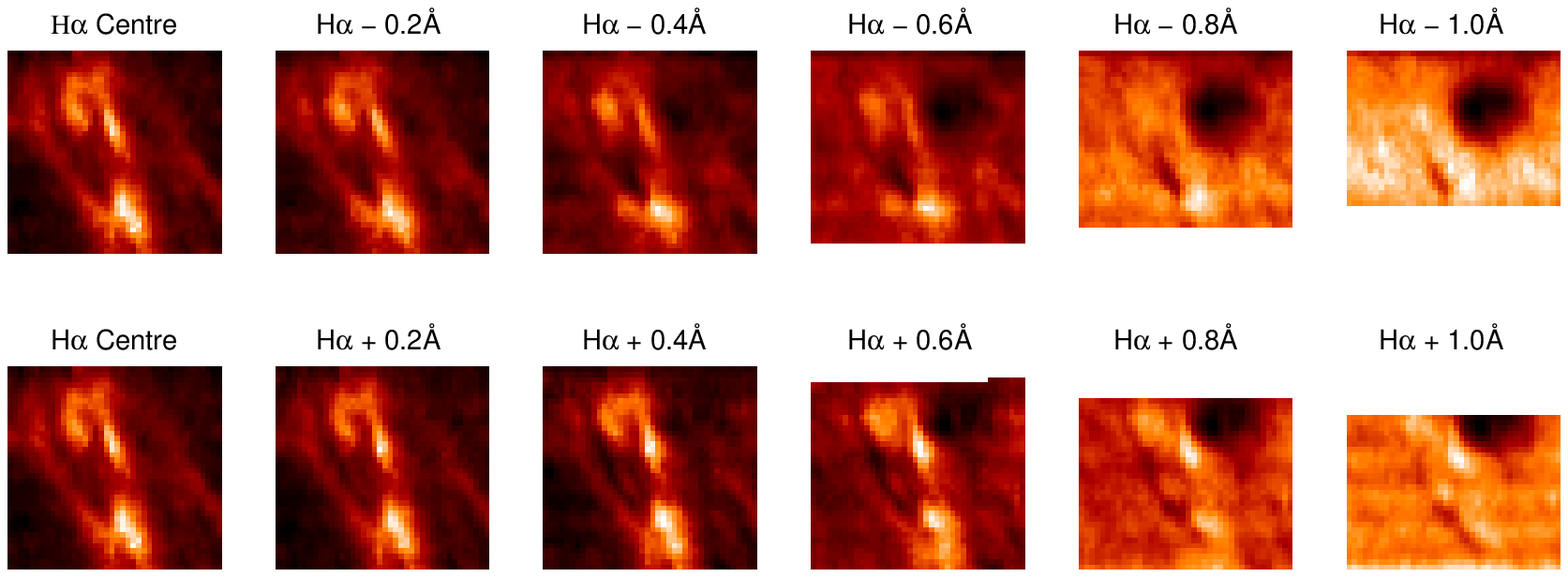}
\end{center}
\vspace{-0.7cm} \caption{C1.2 class solar flare observed in NOAA
10410 active region at 16:03:45 UT on 16 July 2003. Upper panel:
sizes and shapes of the flaring kernels K1, K2 and K3 delimited with
isophote on the level of 75~{\%} of the net maximum brightness of
the kernel. Lower panels: 2D quasi-monochromatic images of the
observed flaring kernels, obtained simultaneously in various
wavelenghts. See main text for details.} \label{fig2}
\end{figure*}

\begin{figure*}[t]
\begin{center}
\includegraphics[width=8.7cm]{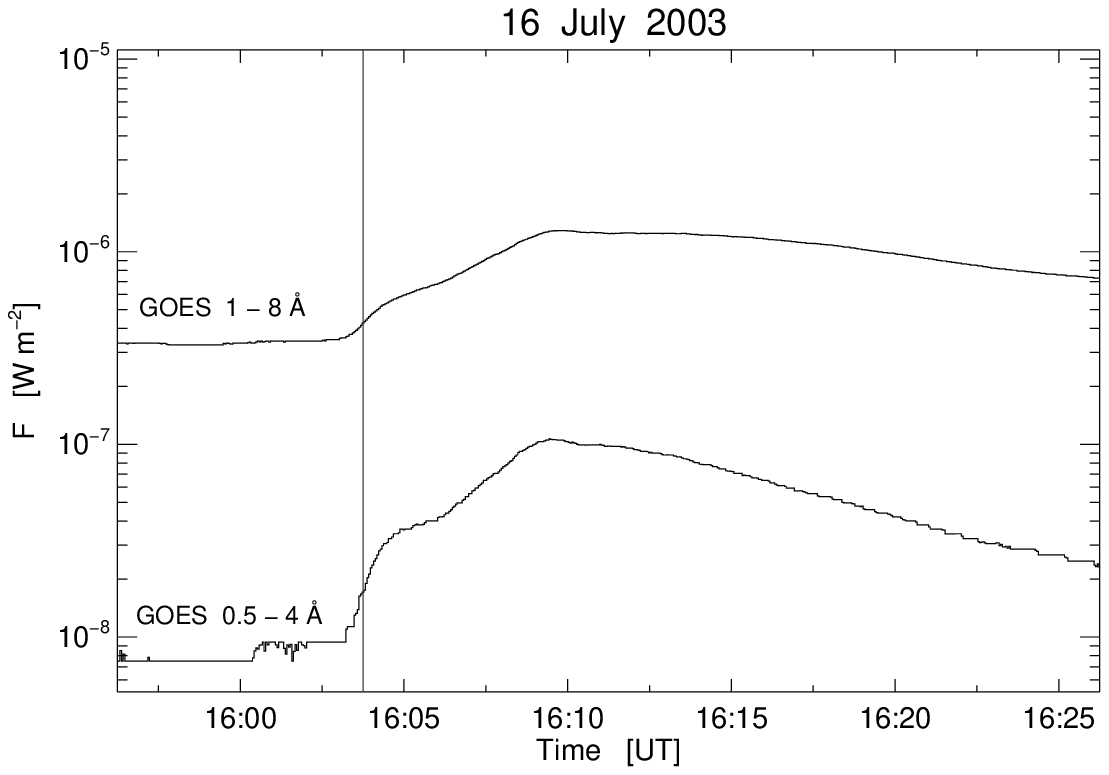}
\includegraphics[width=8.7cm]{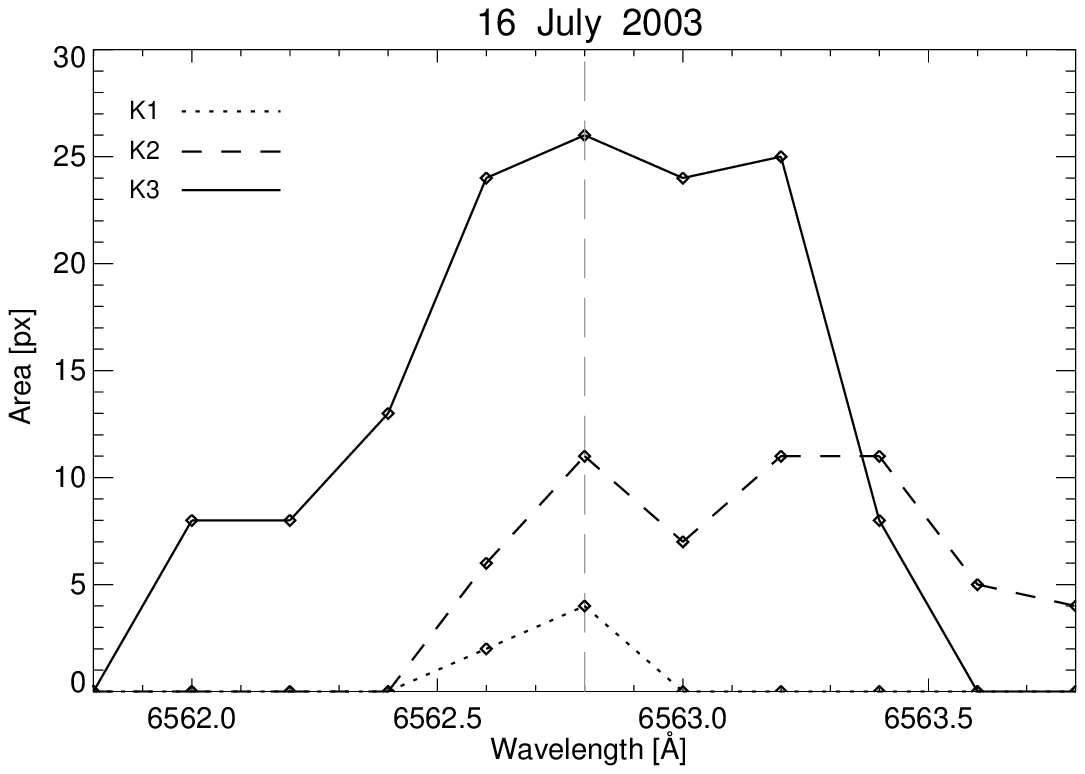}
\end{center}
\vspace{-0.7cm} \caption{Areas of the K1, K2 and K3 flaring kernels
of the C1.2 class solar flare at 16:03:45 UT on 16 July 2003. Left
panel: X-ray \emph{GOES} fluxes in 0.05-0.4 nm and 0.1-0.8 nm bands
recorded during the flare. The vertical line marks 16:03:45 UT.
Right panel: areas of the K1, K2 and K3 flaring kernels (in pixels)
measured simultaneously in various wavelengths in a frame of the
H$\alpha$ line profile (up to $\pm$ 1.0~\AA~from the H$\alpha$ line
center). The vertical dashed line points the H$\alpha$ line center.}
\label{fig021}
\end{figure*}

\begin{figure*}[t]
\begin{center}
\includegraphics[width=13cm]{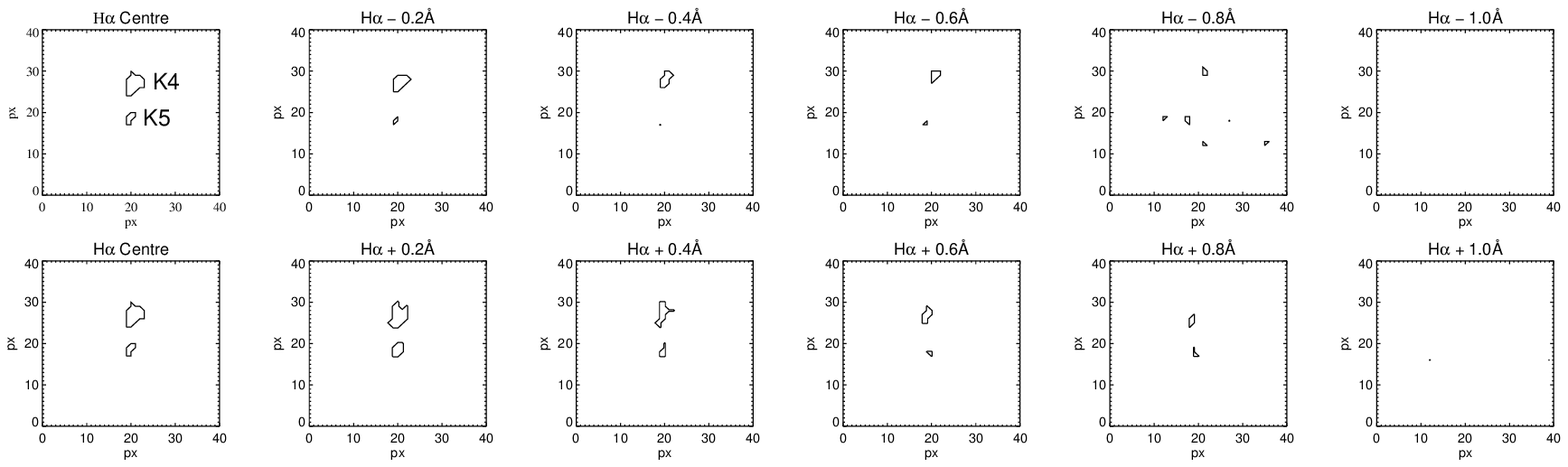}
\includegraphics[width=13cm]{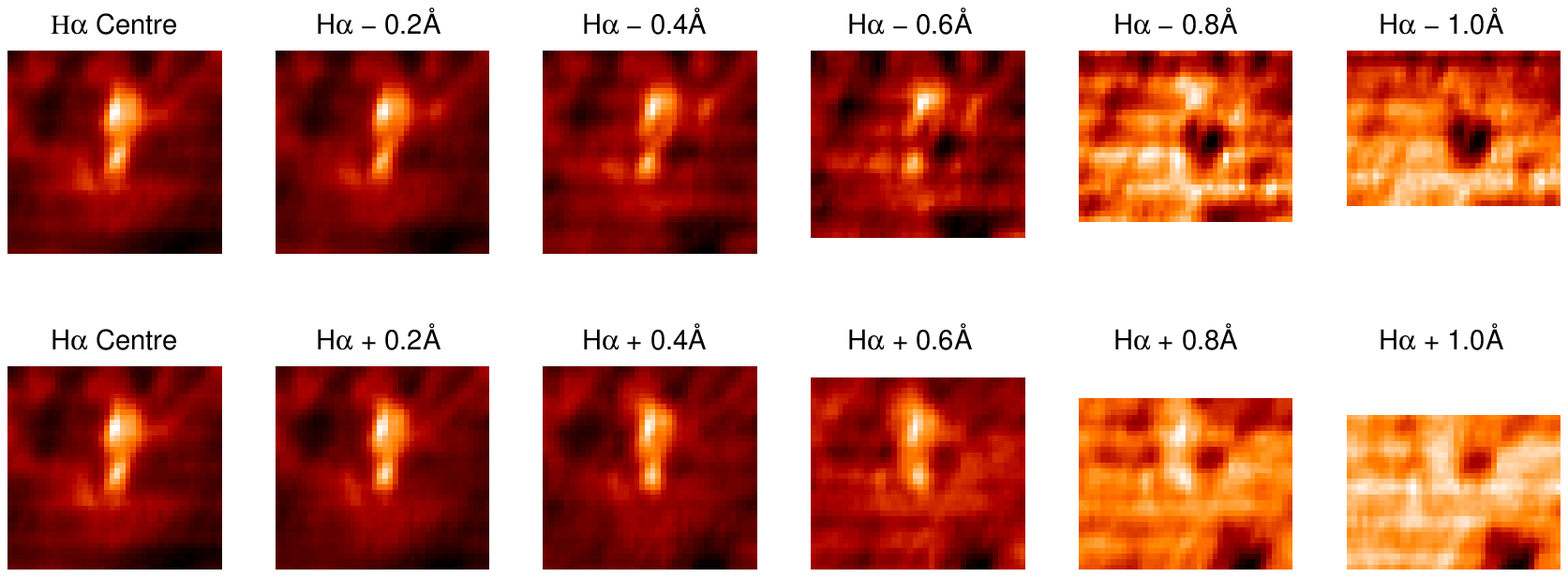}
\end{center}
\caption{B2.5 class solar flare (kernels K4 and K5) in NOAA 10603
active region observed at 07:24:33 UT on 03 May 2004. The upper and
lower panels are the same as on Fig. 1.} \label{fig03}
\end{figure*}

\begin{figure*}[t]
\begin{center}
\includegraphics[width=8.7cm]{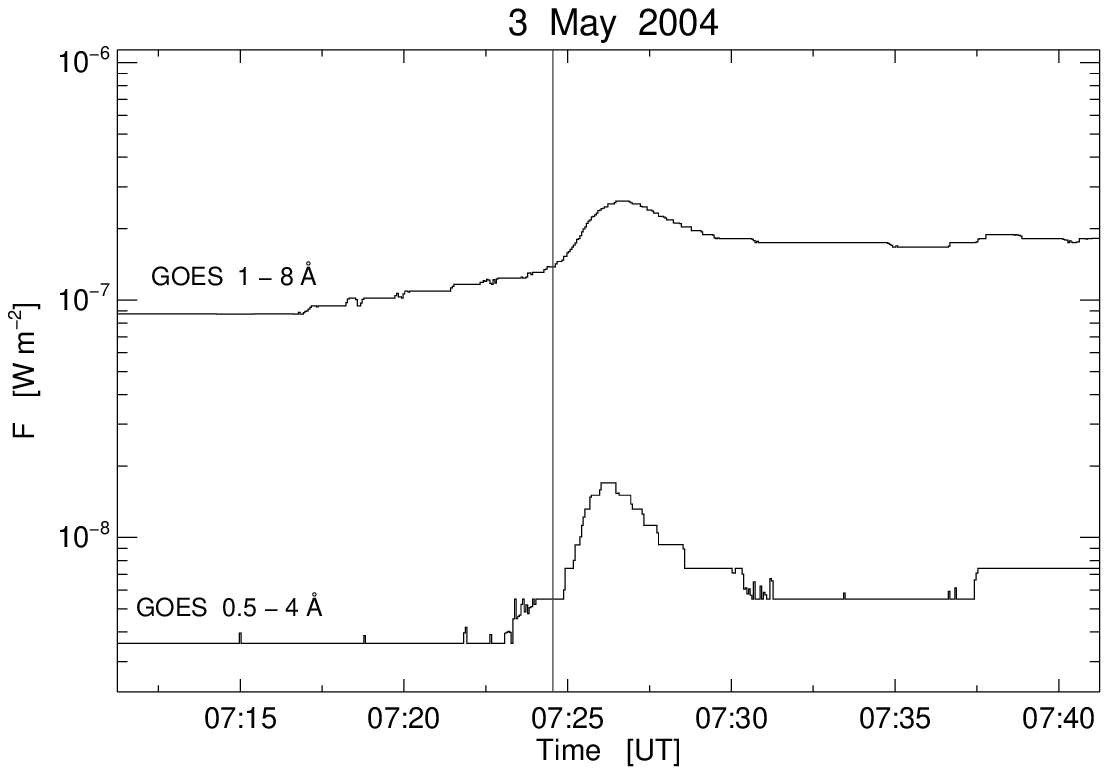}
\includegraphics[width=8.7cm]{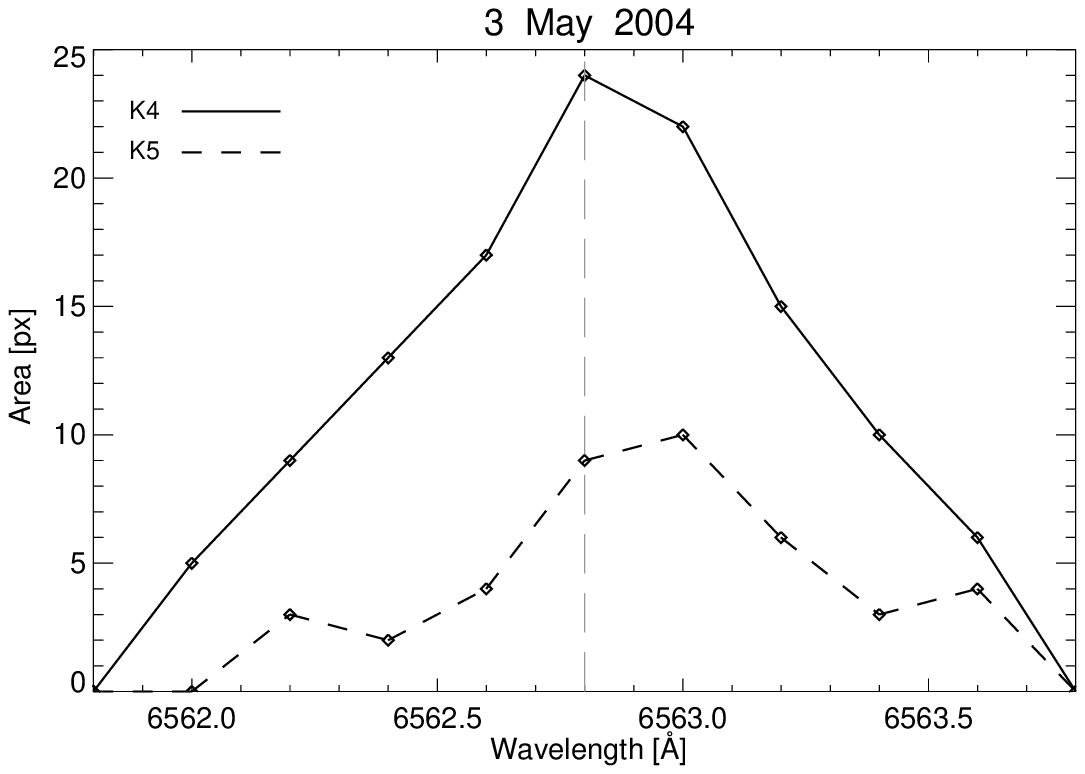}
\end{center}
\vspace{-0.7cm} \caption{Areas of the K4 and K5 flaring kernels of
the B2.5 class solar flare at 07:24:33 UT on 03 May 2004. The left
and right panels are the same as on Fig. 2. } \label{fig021}
\end{figure*}

It is commonly accepted that due to the fast change of the magnetic
$\beta$ parameter between lower corona and photosphere, the
chromospheric parts of the magnetic loops (for example channeling
high energy particle beams exciting chromospheric plasma in course
of the flares), should be roughly cone-shaped (e.g. \citet{gab76};
\citet{fou90}; \citet{stix89}; \citet{asch06}). Unfortunately, the
very strong and variable in time deformations of the apparent shape
of the tiny solar structures caused by highly variable atmospheric
seeing as well as a necessity of the simultaneous 2D observations in
numerous wavelengths made the observational proof of this assumption
very challenging. Taking advantage of unique parameters and
abilities of the \emph{MSDP} spectrograph we made an attempt to
record directly the sizes of the H$\alpha$ flaring kernels
simultaneously in numerous wavelengths of the line profile. While
various parts of the H$\alpha$ line are formed on different depths
in the chromosphere (Vernazza et al. (1973) and (1981)), even taking
into account all obvious factors connected with heterogenous
vertical structure of the plasma, its strong turbulence etc., the
emission recorded in particular part of the line profile could be
used for at least crude determination of the precipitation depth of
the non-thermal electrons as well as for evaluation of the size of
the emitting region.

The observational data and data reduction are described in Section 2
while the results are presented in Section 3 and discussion of the
results in Section 4.

\section{Observations and data reduction}
The H$\alpha$ spectra-imaging observations of the flaring kernels
were collected by us with the Large Coronagraph (\emph{LC}) equipped
with \emph{MSDP} and Solar Eclipse Coronal Imaging System
(\emph{SECIS}) at the Bia{\l}kow Observatory of the University of
Wroc{\l}aw.

The \emph{LC} has a 53 cm diameter main objective, its effective
focal length is equal to 1450 cm. Spatial resolution of the
instrument, usually limited by seeing conditions, is about 1 second
of arc. During windy days the very susceptible to gusts \emph{LC}
was replaced by our Horizontal Telescope with compact Jensch-type
coelostat and main objective of aperture of 15 cm and focal length
of 5 m. The \emph{MSDP} spectrograph has a rectangular (2D) entrance
window, which covers an equivalent area of 325${\times}$41
arcsec$^{2}$ on the Sun (\citet{Mein1991}; \citet{Rompolt1994}) and
creates H$\alpha$ spectra for all pixels inside the field of view
simultaneously.

The \emph{MSDP} spectrograph has a nine-channels \emph{prism-box},
giving (for each pixel of the field of view) nine intensities in the
H$\alpha$ line-profile bandwidth range, separated by 0.4~\AA. Next
the continuous H$\alpha$ profiles are interpolated typically in the
range $\pm$1.2~\AA~from the line center. Due to the basic optical
properties of the \emph{MSDP}-type spectrographs the actual range of
the band-width of the restored profiles depend on the location of
the pixels in the frame of the field of view and thus decrease
slightly toward the edges of the images (this effect is well visible
on Figures 1, 3 and 5). On the basis of the restored H$\alpha$ line
profiles the quasi-monochromatic images were reconstructed in
freely-chosen wavelengths in the line-profile bandwidth range (e.g.
separated in wavelengths by 0.2~\AA~each other as in present work).
The spectra-images created by the spectrograph were recorded with
the fast CCD camera of \emph{SECIS} (512${\times}$512 px$^{2}$,
1~px=1~arcsec$^{2}$, up to 70 images per second)
(\citet{Phillips2000}; \citet{Rudawy2004}). Thus the \emph{MSDP}
spectrograph with \emph{SECIS} is very suitable for studies of the
fast variations of the spectral line profiles emitted by individual
H$\alpha$ flaring kernels or investigations of the solar structures
observed simultaneously in various wavelengths and can be a very
valuable source of the observational data for modeling of the
sub-second flaring processes (\citet{Heinzel2003}).

\begin{figure*}[t]
\begin{center}
\includegraphics[width=13cm]{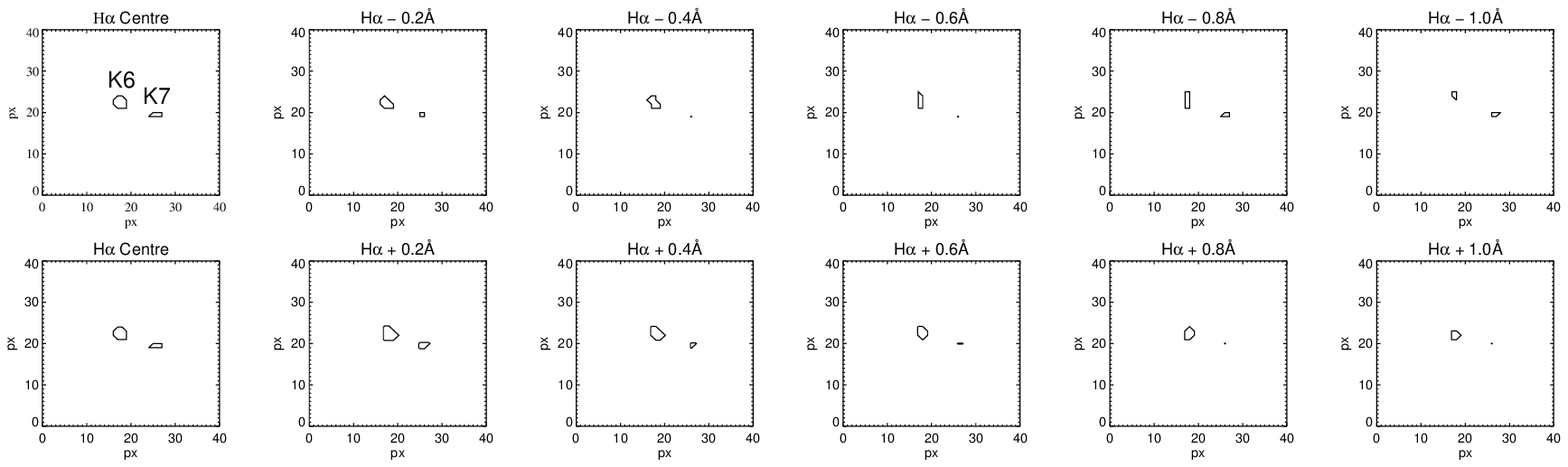}
\includegraphics[width=13cm]{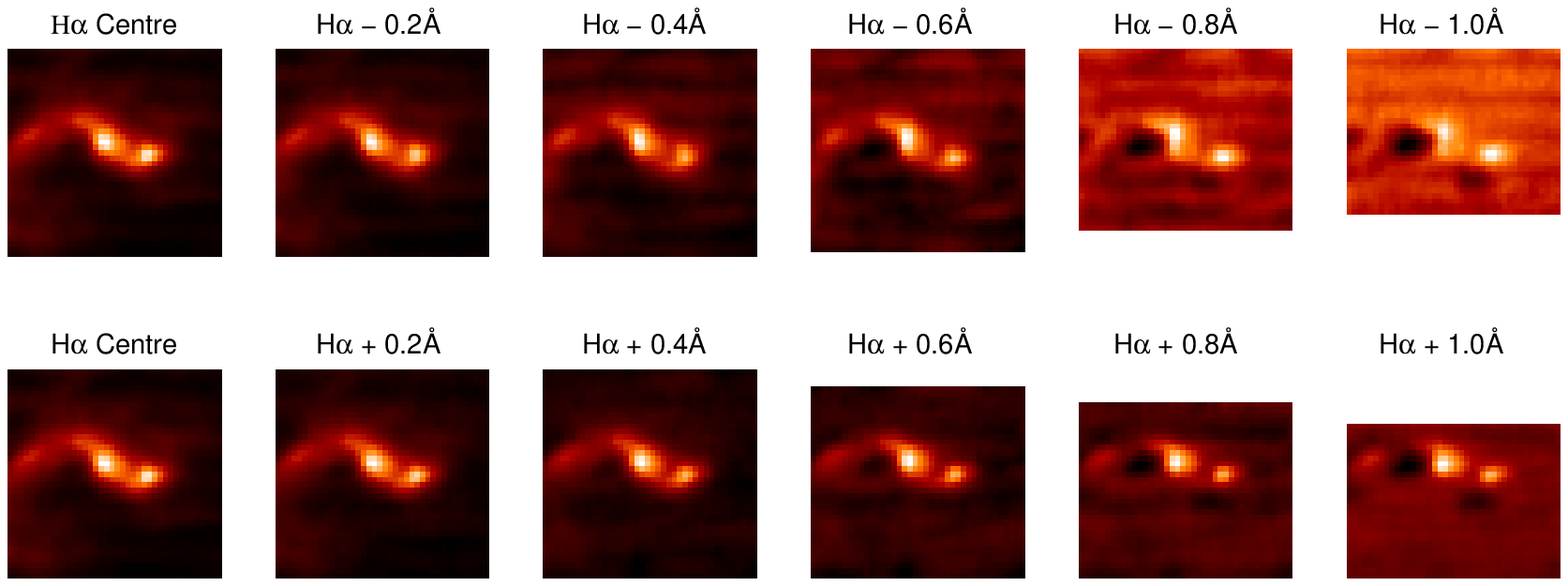}
\end{center}
\caption{C8.3 class solar flare (kernels K6 and K7) in NOAA 10786
active region observed at 08:01:22 UT on 12 July 2005. The upper and
lower panels are the same as on Fig. 1.} \label{fig04}
\end{figure*}

\begin{figure*}[t]
\begin{center}
\includegraphics[width=8.7cm]{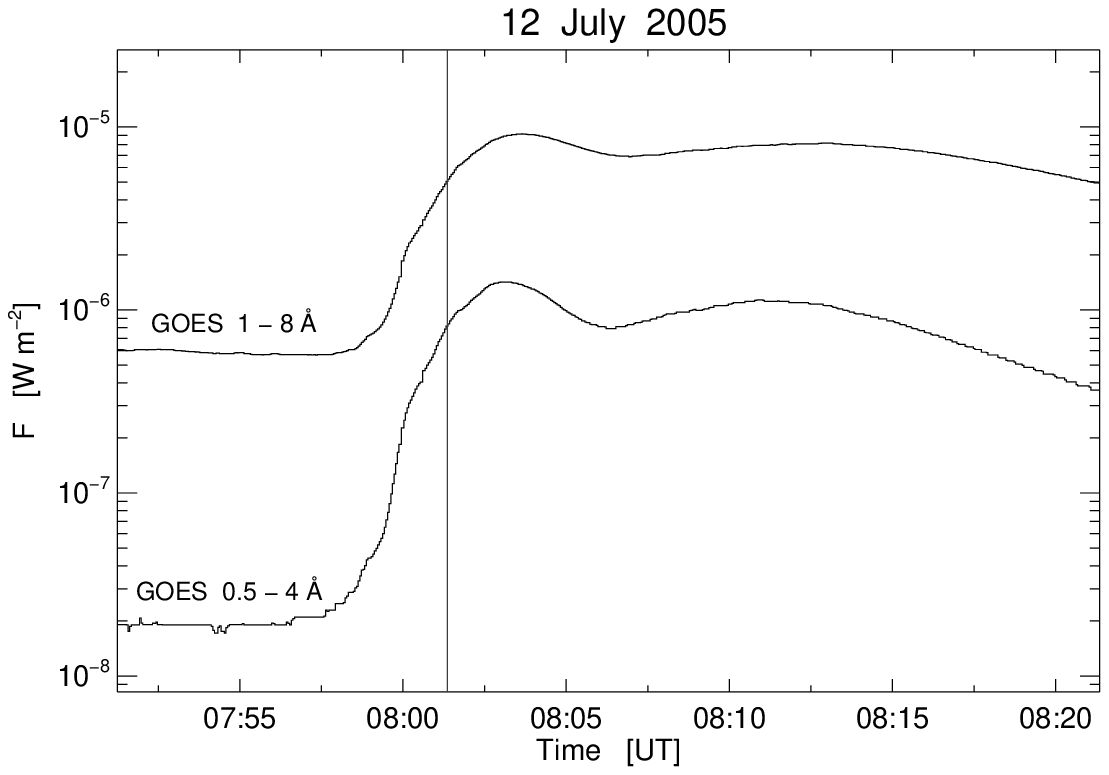}
\includegraphics[width=8.7cm]{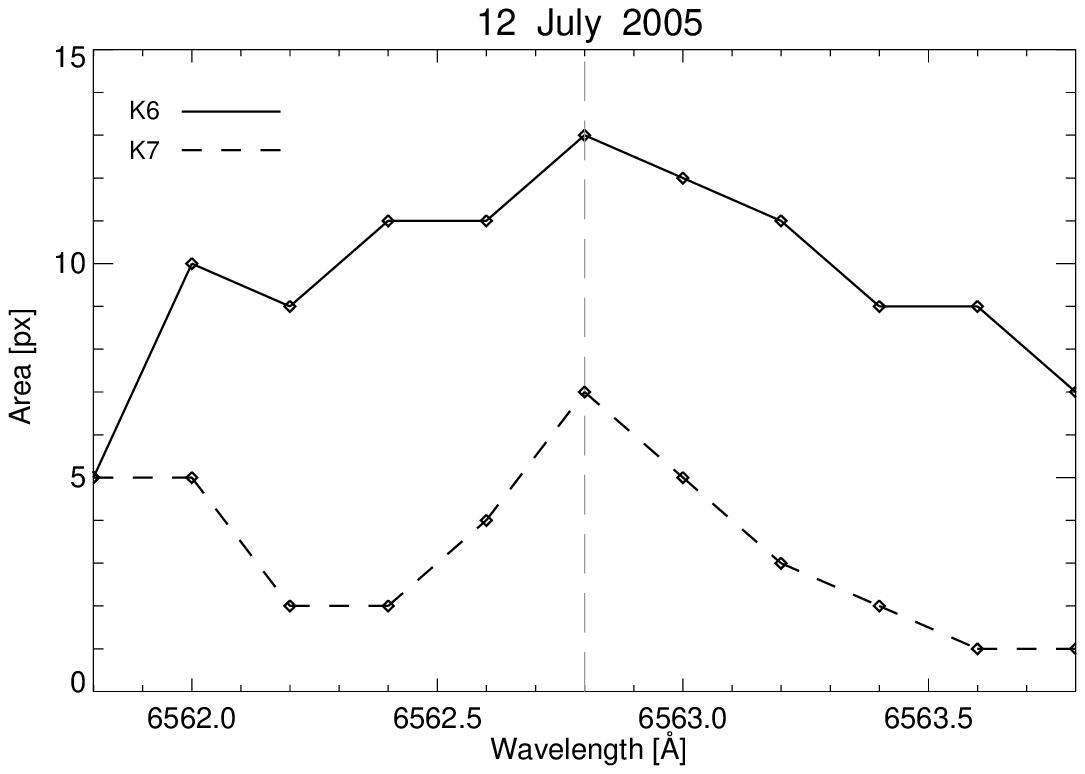}
\end{center}
\vspace{-0.7cm} \caption{Areas of the K6 and K7 flaring kernels of
the C8.3 class solar flare at 08:01:22 UT on 12 July 2005. The left
and right panels are the same as on Fig. 2.} \label{fig021}
\end{figure*}

In a course of three observational seasons 2003-2005 we recorded
data for almost forty solar flares. For each event we recorded 10 or
20 thousand spectra-images with time resolution freely selected
between 0.04 s (25 images per second) and 0.075 s (slightly more
than 13 images per second) – depending on the intensity of the light
beam. More detailed information concerning \emph{MSDP-SECIS} system
and data reduction are available in papers by Radziszewski et al.
(2006) and (2007).

As a result of the numerical reduction of the collected
spectra-images of each particular flare we obtained a series of 10
or 20 thousand of "data-cubes" consisting of thirteen
two-dimensional, quasi-monochromatic images (band-width equal to
0.06 \AA), separated in wavelengths by 0.2~\AA. Due to the actual
localizations of the investigated flaring kernels in the frames of
the fields of views we decided to present the restored profiles
consistently for all flares up to $\pm$1.0~\AA~ from the H$\alpha$
line center only.

The collected data allow us to investigate local temporal variations
of the H$\alpha$ emission and Doppler-shifts of the emitting
material. The temporal variations of the obtained H$\alpha$ data
(profiles, intensities, areas of the emitting kernels etc.) were
compared with temporal variations of the X-ray fluxes recorded with
Reuven Ramaty High Energy Solar Spectroscopic Imager (\emph{RHESSI})
and also with X-ray photometers on boards of the \emph{GOES}
satellites. Our observations, based on high cadence spectra-imaging
spectrograph \emph{MSDP} and \emph{RHESSI} satellite data, have
confirmed the high temporal correlation between HXR and H$\alpha$
emissions for several solar flares (see Radziszewski et al. (2006)
and (2007), and references therein).

From among almost forty events observed with \emph{LC-MSDP-SECIS} or
\emph{HT-MSDP-SECIS} systems we have chosen for this work three
small and medium \emph{GOES} class events, observed from the
beginning of the soft X-ray flux increase and having well separated,
individual flaring kernels. The quality of the selected data was
among the best from all collected (stable pointing of the telescope
as well as good seeing and weather conditions). The selected data
include: C1.2 class solar flare observed in NOAA 10410 active region
at 16:10 UT on 16 July 2003 (Figs. 1 and 2); B2.5 class solar flare
in NOAA 10603 active region observed at 07:26 UT on 03 May 2004
(Figs. 3 and 4) and C8.3 class solar flare in NOAA 10786 active
region observed at 08:03 UT on 12 July 2005 (Figs. 5 and 6).

\section{Results of the observations}

After standard photometrical and geometrical processing of the raw
observational data we established locations, sizes, areas and
brightness of the numerous flaring kernels observed simultaneously
in various wavelengths (in a frame of the H$\alpha$ line profile up
to $\pm$ 1.0~\AA~from the H$\alpha$ line center). It is crucial and
obligatory to stress that all photometrical and geometrical
parameters of the kernels were evaluated strictly at the same moment
for all wavelengths (it means, between another, under exactly the
same influence of the seeing which could be also considered as a
wavelength independent for a whole profile of a single spectral
line). On Figures 1, 3 and 5 there are presented 2D
quasi-monochromatic images of the selected flaring kernels, obtained
simultaneously in various wavelengths and (as a contour maps) sizes
and shapes of the flaring regions delimited with an arbitrary
selected isophote on the level of 75~{\%} of the net maximum
brightness of the kernel (the intensity of the brightest pixel of
the evaluated kernel minus the mean emission of the adjacent quiet
chromosphere). We checked various methods of determination of the
kernels' sizes in various wavelengths. According to us, the applied
method is the most stable and reliable, especially for long series
of data, taken under circumstances of variable atmospheric
transmittance and seeing. We also checked brightness distributions
of the investigated kernels and we did not reveal any local strong
increases of brightness, which could cause false estimations of the
kernel sizes.

\begin{figure}[t]
\begin{center}
\includegraphics[width=6cm]{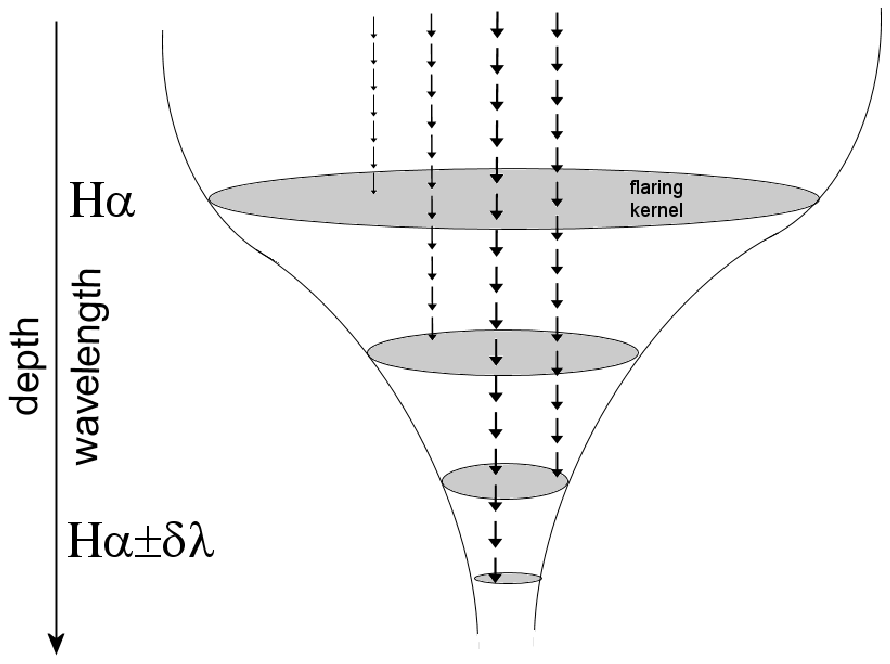}
\end{center}
\caption{Plasma located on various depths in lower part of the
magnetic loops is excited and heated by non-thermal electrons of
various energies. Due to the cone-shaped lower part of the loop, the
actual size of the flaring kernel will depend on the wavelength in a
frame of the H$\alpha$ line profile.} \label{fig05}
\end{figure}

After evaluation of whole series of data (usually covering a period
of 8-10 minutes of the flare) we have found that:

- the most significant changes of emission were detected during the
early phases of the flare, when high-energy particles beams hit the
chromosphere;

- the sizes of the flaring kernels (area or equivalent radius
defined as a radius of the circle having the same area as irregular
surface) of the investigated flares are biggest when measured on
images taken in the H$\alpha$ line center or in wavelengths close to
the line center;

- the effect of the decrease of the kernels' areas in H$\alpha$ line
wings was detected also for series of consecutive images taken
during the increase phases of the flares;

- the sizes of the flaring kernels decreased significantly and
systematically when measured on images taken (at exactly the same
moment) in increased distances from the line center toward the line
wings (see Figures 2, 4 and 6), it means, to some extent, toward the
deeper layers of the chromosphere (see Figure 7). Most of the
kernels show roughly a symmetrical decrease of the visible areas in
increased distances from the line center (kernels K1, K3-K7); one
kernel only (K2) shows a very asymmetrical decrease of the visible
areas (it shrank and disappeared much faster when observed in the
blue wing of the H$\alpha$ line than in the red one). One can also
notice a slight increase of the area of the K7 kernel observed in
the far blue wing of the H$\alpha$ line.

The effect of the decrease of the flaring kernels sizes was
investigated by us only for faint and mid-class solar flares, while
in such flares the individual flaring kernels could be easily
identified and measured.

\section{Discussion}

Taking advantage of the unique parameters and abilities of the
\emph{MSDP} spectrograph we recorded on each spectra-image directly
the sizes of the H$\alpha$ flaring kernels in numerous wavelengths
of the H$\alpha$ line profile simultaneously, with very short
exposure time (0.04-0.075 s) and under exactly the same influence of
the atmospheric seeing. We have found that the sizes of the flaring
kernels are biggest when measured on images taken in the H$\alpha$
line center and decreased significantly when measured on images
taken in increased distances from the line center toward the both
line wings (see Figures 2, 4 and 6, right panels).

Due to the very heterogenous vertical structure and obvious strong
turbulence of the plasma located in the feet of the flaring loop,
fast bulk motions of the plasma caused by chromospheric evaporation
processes it is not straight, time independent and linear dependence
between wavelengths and depth. Nevertheless, our observational
results, obtained for numerous flaring kernels of the faint and
mid-class solar flares could be caused by a cone-shape of the lower
parts of the magnetic loops channeling particle beams exciting
chromospheric plasma (see diagram on Figure 7).

In this work we presented in detail the results obtained for seven
flaring kernels observed in three solar flares only. A brief
inspection of the observations taken during other ten flares of
various \textit{GOES} classes (B and C), recorded during less
favorable seeing conditions and having lower overall quality,
revealed qualitatively the same effect. However, the effect was not
detected by us in strong X3.8 flare observed on 17 January 2005,
having bright and extended flaring structures. Such strong flares
(especially X class events) seem to be not sufficient for our
investigations due to the lack of well separated, individual
kernels.

As a next step of our work we plan to investigate spatial
displacements of the emission arriving co-temporary from the same
kernel but in various wavelengths and temporal changes of kernel
positions (their centroids) observed in selected wavelengths (e.g.
blue wing) of the H$\alpha$ line, as it was already reported by Ji
and co-workers (2004).

\begin{acknowledgements}
This work was supported by the Polish Ministry of Science and Higher Education, grant number N203 022 31/2991.
\end{acknowledgements}


\end{document}